\documentclass[twocolumn,superscriptaddress,floatfix,preprintnumbers, nofootinbib,hyperref]{revtex4-2} 
\usepackage[colorlinks=true,breaklinks=true]{hyperref}
\usepackage[normalem]{ulem}
\usepackage[utf8]{inputenc}
\hypersetup{allcolors=[rgb]{0.0 0.0 0.6},linkcolor=[rgb]{0.75 0.05 0.05}}
\usepackage{amsmath,amssymb}
\usepackage{epsfig} 
\usepackage{graphicx} 
\usepackage{slashed} 
\usepackage{tikz}
\usepackage{color}
\usepackage{tikz-feynman}
\usepackage{url}
\usepackage{array}
\newcolumntype{L}[1]{>{\raggedright\let\newline\\\arraybackslash\hspace{0pt}}m{#1}}
\newcolumntype{C}[1]{>{\centering\let\newline\\\arraybackslash\hspace{0pt}}m{#1}}
\newcolumntype{R}[1]{>{\raggedleft\let\newline\\\arraybackslash\hspace{0pt}}m{#1}}
\usepackage{color}
\usepackage{multirow}
\usepackage{comment}
\usepackage{amssymb}

\DeclareMathOperator{\cm}{cm}
\DeclareMathOperator{\GeV}{GeV}
\DeclareMathOperator{\eV}{eV}

\DeclareMathOperator{\keV}{keV}
\DeclareMathOperator{\MeV}{MeV}

\DeclareMathOperator{\tr}{tr}

\newcommand{\beq}{\begin{equation}}
\newcommand{\eeq}{\end{equation}}

\allowdisplaybreaks

\bibpunct{[}{]}{,}{n}{}{,}

\begin{document}

\title{Glueball dark matter, precisely}

\author{Pierluca Carenza}\email{pierluca.carenza@fysik.su.se}
\affiliation{The Oskar Klein Centre, Department of Physics, Stockholm University, Stockholm 106 91, Sweden
}

\author{Tassia Ferreira}\email{tassia.ferreira@physics.ox.ac.uk}
\affiliation{Department of Physics, University of Oxford, Denys Wilkinson Building, Keble Road, Oxford OX1 3RH, UK
}

\author{Roman Pasechnik}\email{roman.pasechnik@thep.lu.se}
\affiliation{
 Department of Astronomy and Theoretical Physics, Lund University\\
 SE-223 62 Lund, Sweden
}

\author{Zhi-Wei Wang}\email{zhiwei.wang@uestc.edu.cn}
\affiliation{School of Physics, The University of Electronic Science and Technology of China,\\
 88 Tian-run Road, Chengdu, China}

\smallskip

\begin{abstract}
We delve deeper into the potential composition of dark matter as stable scalar glueballs from a confining dark $SU(N)$ gauge theory, focusing on $N=\{3,4,5\}$. To predict the relic abundance of glueballs for the various gauge groups and scenarios of thermalization of the dark gluon gas, we employ a thermal effective theory that accounts for the strong-coupling dynamics in agreement with lattice simulations. We compare our methodology with previous works and discuss the possible sources of discrepancy. The results are encouraging and show that glueballs can account for the totality of dark matter in many unconstrained scenarios with a phase transition scale $20~\MeV\lesssim\Lambda\lesssim10^{10}~\GeV$, thus opening the possibility of exciting future studies.
\end{abstract}

\maketitle

\section{Introduction}

Dark Yang-Mills sectors, which undergo confinement to form stable composite states known as glueballs, may potentially explain the nature of Dark Matter (DM)~\cite{Carenza:2022pjd,Carlson:1992fn,Faraggi:2000pv,Feng:2011ik,Boddy:2014yra,Soni:2016gzf,Kribs:2016cew,Acharya:2017szw,Dienes:2016vei,Soni:2016yes, Soni:2017nlm,Draper:2018tmh,Halverson:2018olu, forestell2017cosmological, Forestell:2016qhc} (see also Ref.~\cite{Das:2018ons}, for a more general class of DM). This type of self-interacting DM has been shown to provide a consistent explanation for the structure of the Universe at small scales and may also help address issues such as the missing satellite problem~\cite{Mateo:1998wg} and the cusp-core problem in the DM distribution at galactic scales~\cite{Spergel:1999mh,de_Blok_2010}.

The first-order confining phase transition at a critical temperature $T_c$, present in these models~\cite{Panero:2009tv,Halverson:2020xpg,Huang:2020crf,Reichert:2021cvs,Kang:2021epo}, makes it highly non-trivial to follow the formation of glueball DM. This challenging and interdisciplinary study requires a detailed knowledge of thermal field theory in a non-perturbative domain, and a productive exchange of results with lattice Quantum Chromodynamics (QCD).

In light of previous studies~\cite{Carenza:2022pjd}, the calculation of glueball relic density is extended to a generic $SU(N)$ gauge group for $N=\{3,4,5\}$ by using a low-energy effective model for the gluon-glueball dynamics~\cite{Sannino:2002wb}. Additionally, different cosmological scenarios are considered to determine the temperature of the dark gluon gas, an important parameter in determining DM glueball abundance. More specifically, the possibility that dark gluons are thermally produced in the primordial plasma or result from a heavy particle (perhaps the inflaton) decay is explored. In a very model-independent way, we determine the glueball models capable of explaining the existence of DM. 

In Sec.~\ref{sec:efflag} we present the effective Lagrangian used to describe the thermal evolution of the dark gluon-glueball plasma. In Sec.~\ref{sec:relic} we discuss how this picture can be merged in a cosmological setting to predict the relic glueball DM abundance. In Sec.~\ref{sec:comp} we analyze the differences between our approach and the ones usually employed in literature, to stress why this approach is the most accurate to date. Sec.~\ref{sec:cosmology} discusses, in a very model-independent fashion, the possible cosmological scenarios in which the dark sector temperature is determined. Finally, in Sec.~\ref{sec:conclusions} we summarize our findings and conclude.

\section{The effective Lagrangian}
\label{sec:efflag}

Studying the confinement-deconfinement phase transition in $SU(N)$ theories requires understanding non-perturbative dynamics. Lattice simulations~\cite{Boyd:1996bx,CP-PACS:1999eop,Scavenius:2001pa,Panero:2009tv}, effective models~\cite{Meisinger:2001cq,Meisinger:2001fi,Dumitru:2000in,Agasian:1993fn,Campbell:1990ak,Simonov:1992bc,Sollfrank:1994du,Carter:1998ti,Drago:2001gd,Renk:2002md,Pisarski:2001pe,KorthalsAltes:1999cp,Dumitru:2001xa,Wirstam:2001ka,Laine:1999hh,Sannino:2002re,Scavenius:2002ru}, and Renormalization Group approaches~\cite{Schaefer:2001cn} have been used to study phase transitions in Yang-Mills theories
effective models. The aim of this work is to use an effective field theory to study the dynamics of the dark gluon-glueball system~\cite{Sannino:2002wb}.

At finite temperature $T$, the $\mathbb{Z}_N$ center of $SU(N)$ is a relevant global symmetry~\cite{Svetitsky:1982gs}, making it possible to construct various gauge invariant operators charged under $\mathbb{Z}_N$. The Polyakov loop, which is charged with respect to the center $\mathbb{Z}_N$ of the $SU(N)$ gauge group (it transforms as $\ell \rightarrow z \ell$ with $z\in \mathbb{Z}_N$), is an example
\begin{equation}
    {\ell}\left(x\right)=\frac{1}{N}{\rm Tr}[{\bf L}]\equiv\frac{1}{N}{\rm Tr}
    \left\{{\cal P}\exp\left[i\,g\int_{0}^{1/T}A_{0}(\tau, \mathbf{x})d\tau\right]\right\}\,,
\end{equation} 
where ${\cal P}$ denotes path ordering, $A_{0}$ is the time component of the vector potential associated with this gauge group, $g$ is the $SU(N)$ coupling constant and $(\tau, \mathbf{x})$ are Euclidean spacetime coordinates.

The Polyakov loop serves as an order parameter for the confinement phase transition in Yang-Mills theory, which occurs at the energy scale $\Lambda$ and is commonly used for this purpose~\cite{Svetitsky:1982gs}. Below the critical temperature, $T_c$, the expectation value of the Polyakov loop operator is zero, while it is non-zero at higher temperatures. The Polyakov Loop Model (PLM) is a mean field approach that models the phase transition in terms of Polyakov loops~\cite{Pisarski:2001pe}. This simplified model captures the essential characteristics of the confinement phase transition in $SU(N)$ theories with $N\geq2$ and has been applied to the study of heavy-ion collisions at the Relativistic Heavy Ion Collider~\cite{Scavenius:2001pa,Scavenius:2002ru}.

The dark gluon-glueball dynamics can be effectively described by considering the dimension-4 glueball field $\mathcal{H}\propto\tr (G^{\mu\nu}G_{\mu\nu})$, with $G^{\mu\nu}$ QCD field strength tensor, and the Polyakov loop $\ell$ in an effective potential given by~\cite{Sannino:2002wb}:
\begin{equation}
    V\left[\mathcal{H},\ell \right]
    =\frac{\mathcal{H}}{2}\ln\left[\frac{\mathcal{H}}{\Lambda^4}\right] + T^4
    {\cal V}\left[ \ell \right]+\mathcal{H}{\cal P}{\left[ \ell \right]} +V_{T}\left[\mathcal{H}\right] \,.
    \label{eq:potential}
\end{equation}
Here, the first term represents the zero-temperature glueball potential, determined by the trace anomaly constraint~\cite{Schechter:1980ak,Schechter:2001ts}. The real polynomials ${\cal V}\left[ \ell \right]$ and ${\cal P}\left[ \ell \right]$ are invariant under $\mathbb{Z}_N$, with coefficients that are fitted to lattice data. The term $V_{T}[\mathcal{H}]$ accounts for thermal corrections, which may involve non-analytic terms in $\mathcal{H}$~\cite{Schaefer:2001cn}.

Remarkably, the potential in Eq.~\eqref{eq:potential} reduces to the glueball dynamics and PLM model in the low and high temperature limits, respectively. Furthermore, the coupling between $\mathcal{H}$ and $\ell$ is the most general interaction term that can be constructed without violating the zero-temperature trace anomaly (see Eq.(21) in Ref.~\cite{Schechter:2001ts}). This approach neglects heavier glueballs and pseudo-scalar glueballs that are described by gauge-invariant operators with different charges under $\mathbb{Z}_N$. Despite its simplicity, this model captures the essential features of the Yang-Mills phase transition.

 \begin{table}[t!]
 	\centering
 	\caption{Parameters for the Polyakov loop potential in Eq.~\eqref{eq:Vell} taken from Ref.~\cite{Huang:2020crf}.}
 	\begin{tabular}{|C{1cm}|C{1cm}|C{1cm}|C{1cm}|
 	}
 		\hline
 		$N$ & 3 & 4 & 5
 		\\
 		\hline
 		$a_0$ & 3.72 & 9.51 & 14.3
 		\\
 		\hline
 		$a_1$ & -5.73 & -8.79 & -14.2 
 		\\
 		\hline
 		$a_2$ & 8.49 & 10.1 & 6.40 
 		\\
 		\hline
 		$a_3$ & -9.29 & -12.2 & 1.74 
 		\\
 		\hline
 		$a_4$ & 0.27 & 0.489 & -10.1 
 		\\
 		\hline
 		$b_3$ & 2.40 & - & -5.61 
 		\\
 		\hline
 		$b_4$ & 4.53 & -2.46 & -10.5 
 		\\
 		\hline
 		$b_6$ & - & 3.23 & -
 		\\
 		\hline	
 	\end{tabular}
 	\label{tab:best-fit}
 \end{table}
 
 \begin{figure*}
    \centering
    \includegraphics[width=0.9\columnwidth]{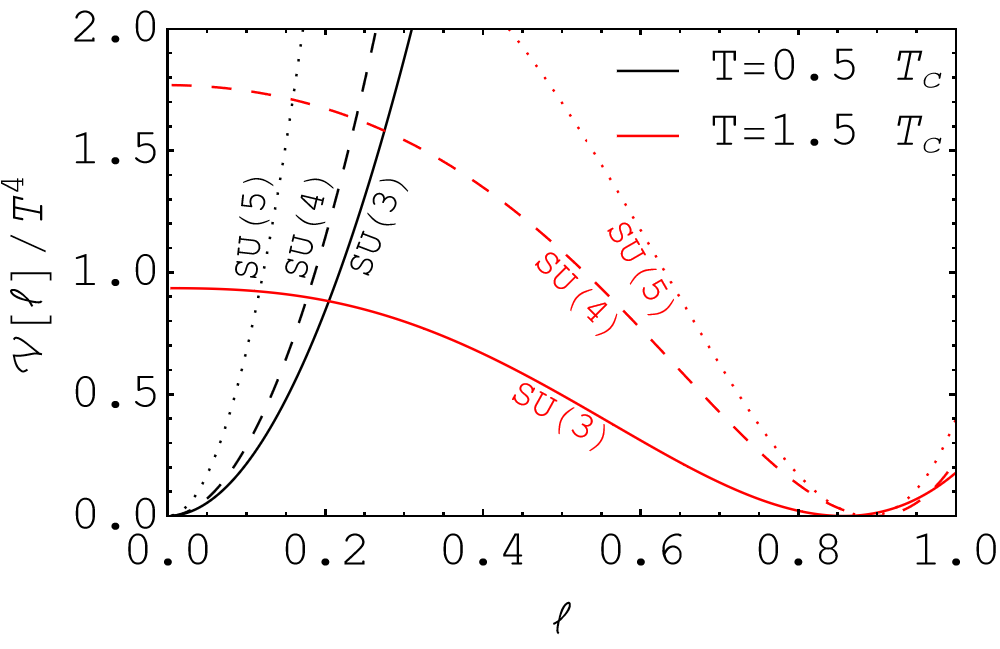}
    \includegraphics[width=0.9\columnwidth]{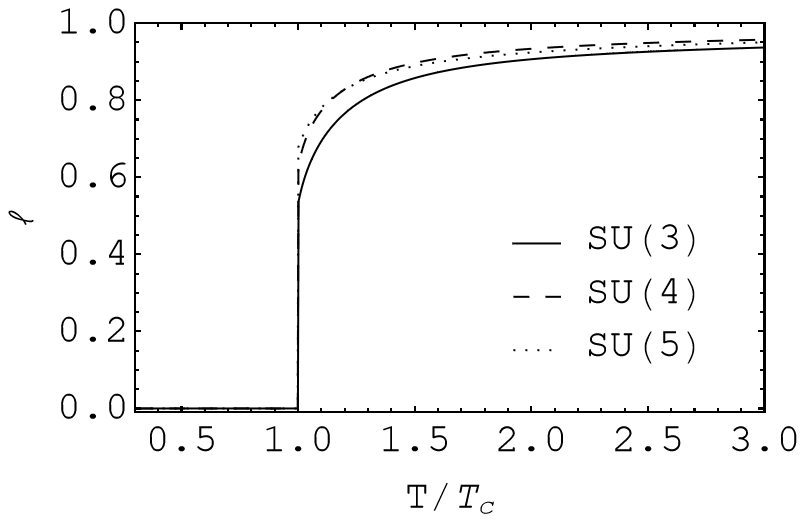}
    \caption{{\it Left panel}: Polyakov loop potential $\mathcal{V}[\ell]$ for different gauge groups: $SU(3)$ solid lines, $SU(4)$ dashed lines and $SU(5)$ dotted lines. The colors correspond to the confined (black) or deconfined (red) phase. Note that the minimum of the potential is arbitrarily set to zero.\\ {\it Right panel}: Polyakov loop evolution as function of the temperature for different gauge groups: $SU(3)$ solid lines, $SU(4)$ dashed lines and $SU(5)$ dotted lines.}
    \label{fig:polyakovN}
\end{figure*}

In the deconfined phase, $T\gg T_{c}$, the PLM term $T^4{\cal V}[\ell]$ dominates, i.~e. dark gluons are the dominant component. The precise relation between the confinement scale $\Lambda$ and the critical temperature of the phase transition $T_{c}$ depends mildly on the gauge group and is determined by lattice simulations. In this paper, we consider $T_{c} =(1.59+1.22/N^{2})\Lambda$ for $N=\{3,4,5\}$~\cite{Lucini:2012wq,Forestell:2016qhc}.

\begin{figure*}
    \centering
    \includegraphics[width=0.9\columnwidth]{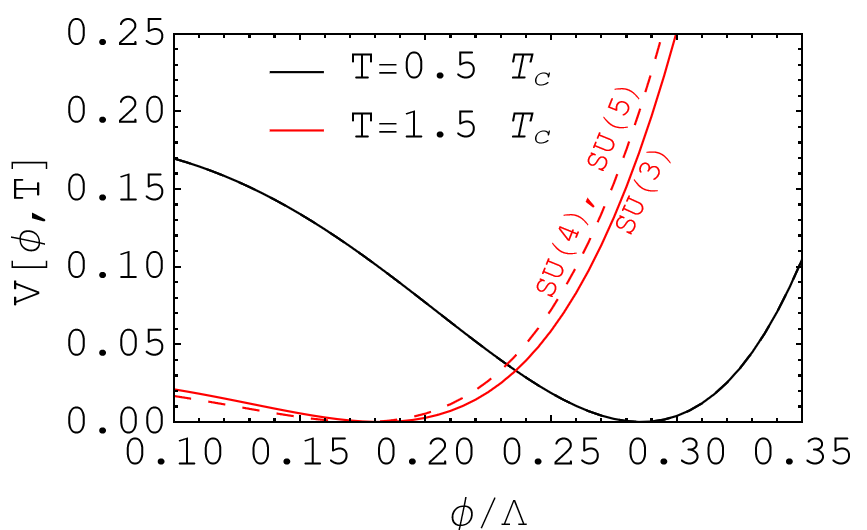} \includegraphics[width=0.9\columnwidth]{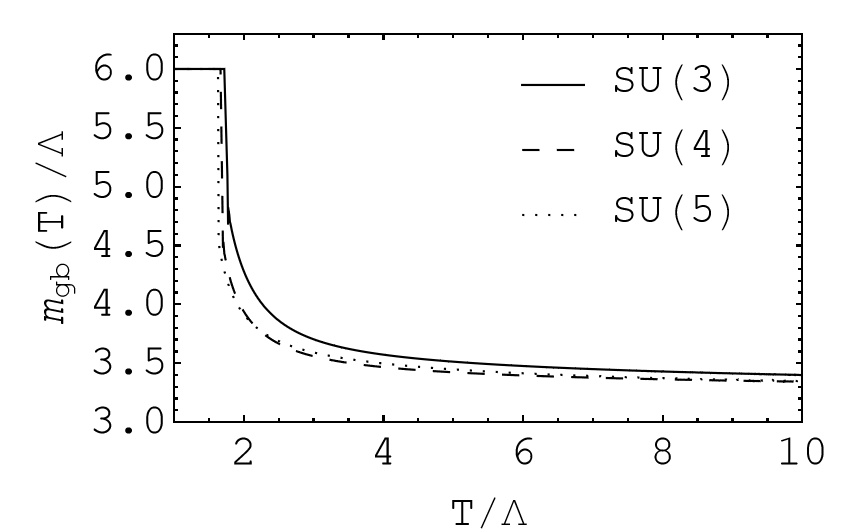}
    \caption{{\it Left panel}: glueball potential $V[\phi,T]$ for different gauge groups: $SU(3)$ solid lines, $SU(4)$ and $SU(5)$ dashed lines. The colors correspond to the confined (black) or deconfined (red) phase. In the confined phase the potential is independent on the gauge group, while it is weakly dependent in the deconfined phase. Note that in this case the potentials for $SU(4)$ and $SU(5)$ are indistinguishable. \\
    {\it Right panel}: effective glueball mass as function of the temperature for various gauge groups: $SU(3)$ solid line, $SU(4)$ dashed line and $SU(5)$ dotted line.}
    \label{fig:glueballpot}
\end{figure*}

The Lagrangian that describes the glueball and Polyakov loop degrees of freedom is given by~\cite{Gomm:1985ut, Ouyed:2001fv, Sannino:2002wb}:
\begin{equation}
    \mathcal{L}=\frac{c}{2}\frac{\partial_{\mu}\mathcal{H}\partial^{\mu}\mathcal{H}}{\mathcal{H}^{3/2}}-V[\mathcal{H},\ell]\,.
\end{equation}
Here, $c=(\Lambda/m_{\rm gb})^{2}/2\sqrt{e}$ is a constant that depends on the glueball mass $m_{\rm gb}$, which we assume to be {$m_{\rm gb}=6\Lambda$}~\cite{Curtin:2022tou}. 
The Polyakov loop is a non-dynamical, homogeneous in space, order parameter that describes the average dynamics of the phase transition. It neglects bubble nucleation, which might have a significant impact on the formation of glueballs, as observed in presence of matter fields~\cite{Asadi:2021pwo,Asadi:2022vkc}, but we leave this discussion for a future work. 
The kinetic term for the glueball field $\mathcal{H}$ is non-standard, due to its dimensionality. To canonically normalize this field, we redefine the glueball field as $\phi$, where $\mathcal{H}=2^{-8}c^{-2}\phi^{4}$, which evolves based on the following effective Lagrangian
\begin{equation}
    \begin{split}
        \mathcal{L}&=\frac{1}{2}\partial_{\mu}\phi\partial^{\mu}\phi-V[\phi,\ell]\,, \\
        V[\phi,\ell]&=\frac{\phi^{4}}{2^{8}c^2}\left[2\ln\left(\frac{\phi}{\Lambda}\right)-4\ln2-\ln c\right]+\\
        &\quad+\frac{\phi^{4}}{2^{8}c^2}\mathcal{P}[\ell]+T^4 {\cal V}\left[ \ell \right]\,,\\
        \mathcal{P}[\ell]&=c_1|\ell|^{2}\,,\\
    \end{split}
    \label{eq:pote}
\end{equation}
where $c_1$ is a free parameter relevant to the determination of the glueball relic abundance. Note that we keep only the lowest order in $\mathcal{P}[\ell]$ satisfying the symmetries. The Polyakov loop potential $\mathcal{V}[\ell]$ for a generic $SU(N)$ gauge group, with $N=\{3,4,5\}$, is determined from symmetry arguments to fit lattice thermodynamic quantities~\cite{Huang:2020crf}
\begin{equation}
    \begin{split}
        \mathcal{V}[\ell]&=T^4\Bigg(-\frac{b_2(T)}{2}|\ell|^2+b_4|\ell|^4+\\
        &-b_3\!\left(\ell^{N}+\ell^{*N}\right)+b_6|\ell|^6 +b_8|\ell|^8 \Bigg)\,,
    \end{split}
    \label{eq:Vell}
\end{equation}
where 
\begin{equation}
    b_2(T)=a_0+a_1\!\left(\frac{T_c}{T}\right)\!+a_2\!\left(\frac{T_c}{T}\right)^{\!2}\!+a_3\!\left(\frac{T_c}{T}\right)^{\!3}\!+a_4\!\left(\frac{T_c}{T}\right)^{\!4}\!\!,
    \label{eq:b2}
\end{equation}
and the parameters of this potential are shown in Tab.~\ref{tab:best-fit}, with the corresponding potentials shown in the left panel of Fig.~\ref{fig:polyakovN}. Here, we notice that the minima of $\ell$ do not differ strongly as a function of the chosen gauge group, albeit the overall potential being quite sensitive to this change. Since the Polyakov loop is a non-dynamical degree of freedom, its temperature evolution is determined by the location of the minimum in the effective potential. Being the order parameter of the phase transition, $\ell$ approaches $1$ at high temperatures and vanishes for temperatures below the critical one. It is possible to numerically find the temperature evolution of $\ell$ by minimizing the potential in Eq.~\eqref{eq:pote} with respect to this variable.
The solution $\ell=0$ denotes the confined phase and it is a global minimum only for temperatures below the critical temperature. In the deconfined phase, the solution $\ell=0$ becomes metastable and a new solution $\ell=\ell_{+}$ becomes the new global minimum. The temperature evolution of the minimum of the Polyakov loop in these potentials is shown in the right panel of Fig.~\ref{fig:polyakovN}. The three gauge groups shown lead to a similar behavior for $\ell$ during the phase transition, with a slightly different critical temperature. Once the minimum of $\ell$ is determined, the Polyakov loop is ``integrated out'' using its equation of motion $\ell=\ell(\phi,T)$, giving rise to a potential for the glueball field in the form $V[\phi,T]=V[\phi,\ell(\phi,T)]$. Moreover, we set the zero-point energy of the glueball field to zero in order to properly describe glueballs as matter. Fig.~\ref{fig:glueballpot}, in the left panel, shows the behavior of the glueball potential as function of the different gauge group. The deconfined phase (red lines) is the only one sensitive to the choice of gauge group, and the dependence is very mild. This is directly correlated with the observation that the minimum of $\ell$ does not change sensibly for different gauge groups with slightly larger $\ell$ when the color increases signaling the stronger jumping and thus a stronger first order phase transition. 
Connected with this potential, it is possible to calculate how the renormalized glueball mass evolves with temperature through the phase transition. This is defined as
\begin{equation}
    m_{\rm gb}^{2}(T)=\frac{\partial^{2}V[\phi,T]}{\partial\phi^{2}}\Bigg|_{\phi=\phi_{\rm min}}\,,
\end{equation}
where $\phi_{\rm min}$ represents the minimum of the glueball field as function of the temperature. This quantity is shown in the right panel of Fig.~\ref{fig:glueballpot}, where we observe a minor difference between the different gauge groups in the deconfined phase. We note that, after confinement, the mass is fixed to be $m_{\rm gb}=6\Lambda$ by construction. The effect of the thermal potential $V_{T}$ will show up in a temperature dependence of the glueball mass in the confined phase. We estimated this effect by using the following potential~\cite{Sannino:2002re}
\begin{equation}
    V_{T}[\phi]=\frac{T^{4}}{2\pi^{2}}\int_{0}^{\infty}dx\,x^{2}\ln\left[1-e^{-\sqrt{x^{2}+\frac{\phi^{2}}{2^{5}c^{2}\,T^{2}}}}\right]\,.
\end{equation}
We find consistent results with Ref.~\cite{Carenza:2022pjd}, since we verified that the impact of this term on the glueball potential is negligible. Namely, the glueball mass is affected less than $10\%$, with comparably small consequences on the considered phenomenology \footnote{With the finite temperature contributions, the minimum of the thermal potential is shifted towards 0 with few percents. The larger effect is on changing the slope around the minimum. This small change cannot be calculated on the relic density because Eq.\eqref{eq:rhoparam} is already an oscillating function and all the uncertainties related to this average process are much bigger than the shift given by the inclusion of $V_T$.}.

By comparing the temperature evolution of the glueball field to lattice simulations~\cite{DElia:2002hkf}, it is possible to impose limitations on the value of the glueball-Polyakov loop coupling $c_{1}$ in Eq.~\eqref{eq:pote}, in the case of $SU(3)$. We found this value to be $c_{1}=1.225\pm0.19$ at 95$\%$ Confidence Level (CL)~\cite{Carenza:2022pjd}.
Since less or no information from the lattice is available for other gauge groups, this will increase the uncertainties on the prediction of the relic abundance. 
To appropriately account for this, we increase the error associated to $SU(3)$ by a factor of $\approx 4$, such that $c_{1}=1.225\pm 0.8$ at 95$\%$ CL for $SU(N)$, \mbox{$N=\{4,5\}$}, and we fix $c_{1}=1.225$ to generate the figures throughout the paper. This parameter significantly affects the location of the minimum of the Polyakov loop. Thus, it will play an important role in determining the initial conditions for the cosmological evolution of the glueball field and on the resulting DM abundance.

\section{Calculation of the glueball relic density}
\label{sec:relic}

As extensively discussed in Ref.~\cite{Carenza:2022pjd}, the glueball field evolves in a Friedmann-Lemaître-Robertson-Walker metric as
\begin{equation}
    \ddot{\phi}+3H\dot{\phi}+\partial_{\phi}V[\phi,T]=0\,,
    \label{eq:original}
\end{equation}
where the dot represents the derivative with respect to the cosmic time $t$ and $H=1/2t$ is the Hubble parameter during a radiation-dominated era. 
The energy density of the glueball field is given by
\begin{equation}
    \rho=\frac{1}{2}(\dot{\phi})^{2}+V[\phi,T]\,,
\end{equation}
and this quantity is used to compute the glueball DM relic density. During a radiation-dominated era, where we expect the confinement to happen, there is a relation between the time and photon temperature $T_{\gamma}$
\begin{equation}
    t=\frac{1}{2}\sqrt{\frac{45}{4\pi^{3}g_{*,\rho}(T_{\gamma})}}\frac{m_{P}}{T_{\gamma}^{2}}\,,
\end{equation}
where $m_{P}=1.22\times10^{19}$~GeV is the Planck mass and $g_{*,\rho}$ is the number of energy-related degrees of freedom.
Thus, Eq.~\eqref{eq:original} can be rewritten as a function of the temperature as
\begin{equation}
    \frac{4\pi^{3}g_{*,\rho}}{45m_{P}^{2}}\xi_T^4 T^{6}\frac{d^{2}\phi}{dT^{2}}+\frac{2\pi^{3}}{45m_{P}^{2}}\frac{dg_{*,\rho}}{dT}\xi_T^4T^{6}\frac{d\phi}{dT}+\partial_{\phi}V[\phi,T]=0\,,
    \label{eq:KG2}
\end{equation}
where the temperature of the dark sector $T$, such that $T_{\gamma}/T=\zeta_{T}$, governs the moment of the phase transition and $\zeta_{T}$ is model dependent, being determined by the interactions with the visible sector. 
Moreover, the second term can be neglected for a large range of temperatures as $g_{*,\rho}$ is constant except at a few isolated events of entropy production (the QCD phase transition, for example). We consider it as a free parameter, depending on the moment in which the phase transition happens and, if $g_{*,\rho}$ is constant, it can be reabsorbed in the definition of the temperature ratio by defining $\xi_{T}'=\xi_{T}g_{*,\rho}^{1/4}$. 

\begin{figure*}
    \centering
    \includegraphics[width=0.9\columnwidth]{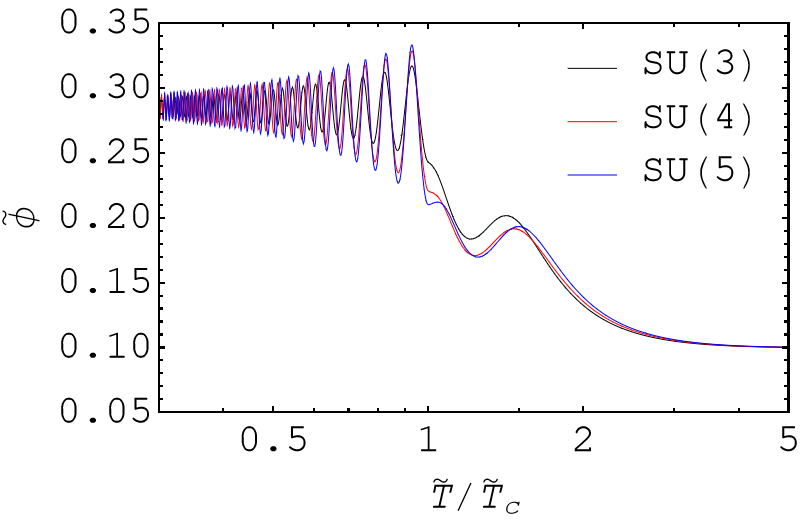}
    \includegraphics[width=0.9\columnwidth]{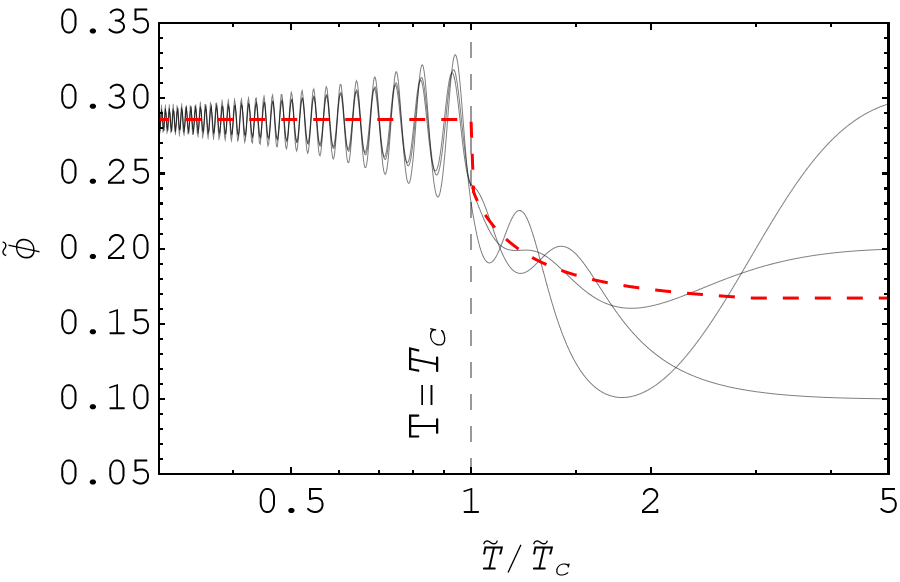}
    \caption{{\it Left panel}: glueball evolution obtained by solving Eq.~\eqref{eq:KGrescaled} for $SU(3)$ (black line), $SU(4)$ (red line) and $SU(5)$ (blue line) with $\mu=0.05$ and initial condition $\tilde{\phi}(\tilde{T}_{i})=0.1$.\\
  {\it Right panel}: similar to the left panel but only for $SU(3)$, with $\mu=0.05$ and different initial conditions. The vertical dashed line marks the phase transition, and the red dashed line shows the evolution of the minimum of
the glueball potential.}
    \label{fig:phivsT}
\end{figure*}

The glueball evolution is analogous to a damped oscillator in a non-linear potential, and the energy stored in these oscillations around $\phi_{\rm min} \approx 0.28\Lambda$ 
\begin{equation}
    \rho=\frac{2\pi^{3}g_{*,\rho}}{45m_{P}^{2}}T^{6}\xi_{T}^{4}\left(\frac{d\phi}{dT}\right)^{2}+V[\phi,T]\,,
\end{equation}
will determine the relic DM abundance since this energy density scales as $\sim T^{3}$, as Cold DM (CDM), when the harmonic approximation is valid. 

Using the following definitions $\phi=\Lambda^{4}\tilde{\phi}$, $\rho=\Lambda^{4}\tilde{\rho}$, $V=\Lambda^{4}\tilde{V}$, $\xi_{T}'=\xi_{T}g_{*,\rho}^{1/4}$, $T=\Lambda \tilde{T}$ and $\mu^{2}=4\pi^{3}\xi_{T}'^{4}\Lambda^{2}/45m_{P}^{2}$, Eq.~\eqref{eq:KG2} can be written as
\begin{equation}
    \begin{split}
        & \mu^{2} \tilde{T}^{6}\frac{d^{2}\tilde{\phi}}{d\tilde{T}^{2}}+\partial_{\tilde{\phi}}\tilde{V}[\tilde{\phi},\tilde{T}]=0\,,\\
        & \tilde{\rho}=\frac{\mu^{2} \tilde{T}^{6}}{2}\left(\frac{d\tilde{\phi}}{d\tilde{T}}\right)^{2}+\tilde{V}[\tilde{\phi},\tilde{T}]\,,
    \end{split}
    \label{eq:KGrescaled}
\end{equation}
which is solved from an arbitrary temperature $T_{i}>T_{c}$ down to some final temperature $T_{f}$ in the confined phase, and from this temperature on the evolution is simply determined by the cosmological expansion. In Fig.~\ref{fig:phivsT} we show the results of Eq.~\eqref{eq:KGrescaled}. In the left panel we show the evolution of the glueball field for different gauge groups. The amplitude of the oscillations in the confined phase is related to the relic energy density. Therefore, we expect that, for the choice of parameters shown in the figure, $SU(5)$ gives an abundance slightly larger than $SU(4)$ and significantly larger than $SU(3)$. In the right panel of Fig.~\ref{fig:phivsT} we focus only on the case of $SU(3)$, to highlight the dependence on the initial conditions. the evolution of the glueball field is shown for three different choices of initial conditions in the hot phase (on the right of the vertical dashed line denoting the critical temperature). After the phase transition the evolution of the three lines is qualitatively similar, suggesting that there is a weak dependence on the initial conditions. 
This was already observed in Ref.~\cite{Carenza:2022pjd}. Intuitively, in the hot phase, the glueball field is set to some arbitrary initial condition then starts to roll towards the minimum of its potential. This happens when the effective glueball mass $ m_{\rm gb}\simeq 3.5\Lambda$ in the deconfined phase becomes larger than the Hubble parameter, in the opposite case the glueball field evolution is frozen.
Starting from the moment in which the Hubble parameter is comparable with the glueball mass, the glueball field efficiently converges to the minimum of its potential. Given the discontinuous nature of the phase transition, the minimum of the glueball potential jumps from the value immediately before the confinement to evolve in a temperature-independent potential. It means that regardless of the evolution of the glueball before the phase transition, only the initial condition set by the properties of the glueball potential $V[\phi,T]$ in the deconfined phase is relevant to predict the glueball relic density. Moreover, the velocity of the field can be taken to be equal to zero since any velocity acquired immediately after the confinement is considerably larger than the velocity accumulated in the evolution in the deconfined phase.
This feature can be interpreted as a washing out of the initial conditions due to the strong first order phase transition.

\begin{figure}
    \centering
    \includegraphics[width=0.9\columnwidth]{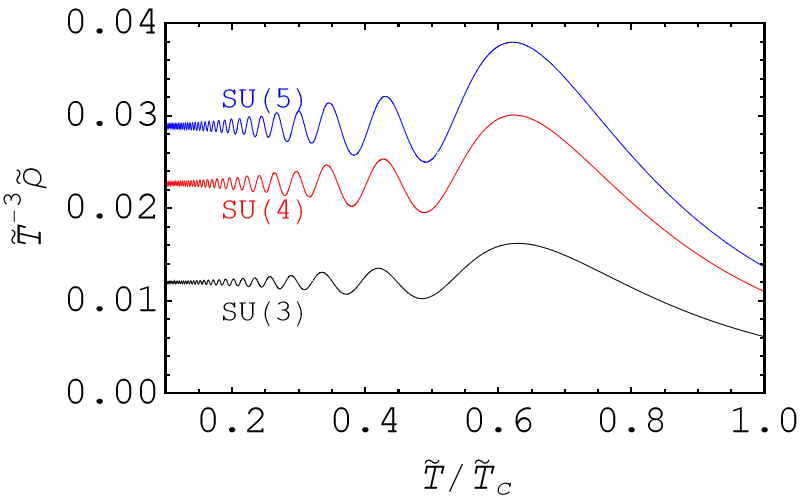}
    \caption{Evolution of the glueball energy density obtained by solving Eq.~\eqref{eq:KGrescaled} for $SU(3)$ (black line), $SU(4)$ (red line) and $SU(5)$ (blue line) with $\mu=1$. The lines start in the point where the phase transition happens, which is different for the three cases shown here.}
    \label{fig:rhoT}
\end{figure}

We discovered that, although the numerical solution of Eq.~\eqref{eq:KGrescaled} is exact, the temperature-dependent potential makes its evaluation computationally expensive. Thanks to the weak sensitivity on the initial conditions, a good approximation is given by solving Eq.~\eqref{eq:KGrescaled} only in the confined phase. The main advantage is using a temperature-independent potential to evolve the glueball field from the critical temperature $T_{c}$ down to a final temperature, $T_{f}$, taking as initial conditions for the glueball field its minimum value $\tilde{\phi}_{\rm min}$ just before the phase transition (at a temperature $\tilde{T}_{c}+\epsilon$ with $\epsilon>0$) and a vanishing first derivative 
\begin{equation}
    \begin{cases}
        & \mu^{2} \tilde{T}^{6}\frac{d^{2}\tilde{\phi}}{d\tilde{T}^{2}}+\partial_{\tilde{\phi}}\tilde{V}[\tilde{\phi}]=0\,,\\
        \vspace{0.1cm}\\
        &\tilde{\phi}(\tilde{T}_{c})=\tilde{\phi}_{\rm min}(\tilde{T}_{c}+\epsilon) \,,\\
        \vspace{0.1cm}\\
        &\frac{d\tilde{\phi}}{d\tilde{T}}(\tilde{T}_{c})=0\,,
    \end{cases}
    \label{eq:KGrescaled0}
\end{equation}
where the potential and the energy density are
\begin{equation}
    \begin{split}
        & \tilde{\rho}=\frac{\mu^{2} \tilde{T}^{6}}{2}\left(\frac{d\tilde{\phi}}{d\tilde{T}}\right)^{2}+\tilde{V}[\tilde{\phi}]\,,\\
        \tilde{V}[\tilde{\phi}]&=\frac{\tilde{\phi}^{4}}{2^{8}c^{2}}\left(2\ln\tilde{\phi}-4\ln2-\ln c\right)+\frac{1}{2e}\,.
    \end{split}
\end{equation}
The energy density obtained from this calculation is shown in Fig.~\ref{fig:rhoT} for three different gauge groups. It is clear that, as the temperature drops, glueballs behave like CDM and their relic density is redshifted as $\sim a^{-3}$, where $a\sim 1/T$ is the scale factor. Note that for $T/T_{c}\gtrsim 0.2$ the glueball energy density does not redshift as CDM because in this intermediate regime non-linearities in the potential lead to deviation from the perfect fluid behavior for the glueball field, i.e. the glueball field does not behave like a pressureless fluid, $p\neq 0$. Indeed, only for $T/T_{c}\lesssim 0.2$, when the glueball field oscillate around its minimum under the influence of an effectively quadratic potential, the perfect fluid approximation is valid. Thus, the quantity shon in Fig.~\ref{fig:rhoT} approaches asymptotically a constant value. One can picture the intermediate phase $0.2\lesssim T/T_{c}\lesssim 1$ as a continuation of the phase transition, when the non-relativistic glueball population is still being established. This result confirms the expectation from Fig.~\ref{fig:phivsT} that the relic density with $SU(5)$ is larger than the one obtained with the two other gauge groups.

Since the energy density is an oscillating quantity, we evaluate an average during the last oscillations before $\tilde{T}_{f}$
\begin{equation}
   \left \langle\frac{\tilde{\rho}}{\tilde{T}^{3}}\right\rangle_{f}=\frac{1}{0.3\tilde{T}_{f}}\int_{\tilde{T}_{f}}^{1.3\tilde{T}_{f}}\frac{\tilde{\rho}(\tau)}{\tau^{3}}d\tau\,,
   \label{eq:rhoparam}
\end{equation}
and this quantity saturates to a value independent on the phase transition scale.
Then the relic density today is calculated by diluting the energy density in Eq.~\eqref{eq:rhoparam} with a factor $(T_{\gamma,0}/\zeta_{T}T_{f})^{3}$, to take into account the Universe expansion as
\begin{equation}
\begin{split}
    \Omega_{g}h^{2}&=\frac{\Lambda}{\rho_{c}/h^{2}}\left \langle\frac{\tilde{\rho}}{\tilde{T}^{3}}\right\rangle_{f}T_{f}^{3}\left(\frac{T_{\gamma,0}}{\zeta_{T}T_{f}}\right)^{3} \\
    &=0.12\zeta_{T}^{-3}\frac{\Lambda}{\Lambda_{0}}\,,
    \label{eq:lambda0}
\end{split}
\end{equation}
where the critical density is $\rho_{c}/h^{2}=1.05\times10^{4}\eV\cm^{-3}$, 
$h=0.674$ and the temperature of the photon bath today is $T_{\gamma,0}=0.235$~meV~\cite{ParticleDataGroup:2022pth}. We defined $\Lambda_{0}$ as the phase transition scale that makes glueballs become the totality of DM. Naively, after combining $T_{\gamma,0}^3/\left(\rho_c/h^2\right)$, we would roughly have $\Lambda_0\sim \langle\frac{\tilde{\rho}}{\tilde{T}^{3}}\rangle_{f}^{-1}\rm{eV}$. Note that here we introduce the $\zeta_{T}$ parameter, related to the glueball and photon temperatures, and not $\zeta_{T}'$ which also includes the number of degrees of freedom in the Universe. This implies that the dependence on this parameter, equivalently $\mu$, is weak in the limit $\mu\ll1$, realized in the relevant case when the phase transition scale is much lower than the Planck scale~\footnote{The glueball energy density calculated by means of Eq.~\eqref{eq:KGrescaled0}, for sufficiently small values of $\mu$, becomes independent on this parameter. This is precisely the same behavior described in Ref.~\cite{Carenza:2022pjd} before Eq.~11. Intuitively, if the glueball self-interactions are too weak (a large $\Lambda$), at the moment of their formation the cannibalism is not efficient compared to the Universe expansion, a situation never realized in our applications.}. 
This approximation is proven to be excellent, predicting the relic density with less than $\sim10\%$ uncertainty compared to the exact result. The good agreement of the two results is a proof that the detailed behavior of the glueball field in the deconfined phase has a minimal impact on the relic density prediction.

\section{Results and comparison with literature}
\label{sec:comp}

In Tab.~\ref{tab:relic} we summarize our findings for the glueball relic density. For each gauge group $SU(N)$ considered, labeled by $N=\{3,4,5\}$ (first column), we recall the range of variability for the term $c_{1}$ at $95\%$ CL (second column). Then we show the results of Eq.~\eqref{eq:rhoparam} (third column) and the corresponding $\Lambda_{0}$ (fourth column), as defined in Eq.~\eqref{eq:lambda0}, for a calculation running down to $\tilde{T}_{f}=0.1$. 

Note that the value of $\Lambda_{0}$ for $N=3$ is $20\%$ larger compared to the one presented in Ref.~\cite{Carenza:2022pjd} because of the different critical temperatures considered.
The values of the relic density found in Tab.~\ref{tab:relic} differ strongly from the estimates in literature, since they report a relic density one or two orders of magnitude higher~\cite{Carlson:1992fn,Halverson:2016nfq,Forestell:2016qhc}. 

The reason for this difference is due to a combination of several effects: inclusion of the higher-order interactions leading to $n\to m$ transitions; energy budget of the dark gluon field partially used for bubble formation; and different equation of state for the glueball field immediately after the phase transition.

In the following we expand each point individually. First, when solving the evolution equation, we are considering a non-trivial potential for the glueball field that, in the confined phase, can be expanded around the minimum $\tilde{\phi}_{\rm min}=4e^{-1/4}\sqrt{c}$ as
\begin{equation}
    \begin{split}
        \tilde{V}[\tilde{\phi}]&=\frac{\tilde{\phi}^{4}}{2^{8}c^{2}}\left(2\ln\tilde{\phi}-4\ln2-\ln c\right)+\frac{1}{2e}\simeq\\
        &\simeq\frac{1}{4c\,e^{1/2}}\delta\tilde{\phi}^{2}+\frac{5}{48c^{3/2}e^{1/4}}\delta\tilde{\phi}^{3}+\\
        &\quad\quad+\frac{11}{768c^{2}}\delta\tilde{\phi}^{4}+\frac{e^{1/4}}{2560c^{5/2}}\delta\tilde{\phi}^{5}+...\,,
    \end{split}
    \label{eq:expansion}
\end{equation}
where $\delta\tilde{\phi}=\tilde{\phi}-\tilde{\phi}_{\rm min}$. Only for $\delta\tilde{\phi}\ll1$ the perturbative concept of particle is valid.

\begin{table}[t!]
    \centering
    \begin{tabular}{|C{1cm}|C{2cm}|C{2cm}|C{2cm}|}
    \hline
    $N$ & $c_{1}$ & $100\times\left \langle\frac{\tilde{\rho}}{\tilde{T}^{3}}\right\rangle_{f}$ &$\Lambda_{0}~(\eV)$ \\
    \hline
    3 &$1.225\pm0.19$ &$0.59\substack{+0.15\\ -0.14}$ & $133\pm32$\\
    4 &$1.225\pm0.8$ & $1.1\substack{+1.0 \\ -0.9}$& $204\pm168$\\
    5 &$1.225\pm0.8$ &$1.3\substack{+1.2 \\ -1.0}$& $139\pm109$ \\
    \hline
    \end{tabular}
    \caption{Results of the calculation of the relic density. The first column represents the gauge group, the second one is the value of $c_{1}$ at $95\%$ CL used in the calculation of the third column, which also shows the result of Eq.~\eqref{eq:rhoparam} at $95\%$ CL, evaluated for $\mu=10^{-3}$ and $\tilde{T}_{f}=0.1$. With these values it is possible to calculate the glueball relic density from Eq.~\eqref{eq:lambda0} and the range of $\Lambda_{0}$ is reported in the fourth column.}
    \label{tab:relic}
\end{table}

However, our calculation is always valid, including all the interactions predicted by the glueball potential. It can be perturbatively understood as including all the possible interactions corresponding to different powers of the expansion in Eq.~\eqref{eq:expansion}. In the literature, it is usually considered that glueballs interact only with a $\phi^5$ interaction, which makes the $3\to2$ annihilation the only relevant process for DM formation. By contrast, in our case also the lower order terms are included.

\begin{figure}

\begin{tikzpicture}[scale=1., transform shape]
\begin{feynman}
\vertex (a) {\(\phi_{i}\)} ;
\vertex [above=0.7 cm of a] (b) {\(\phi_{i}\)};
\vertex [below=0.7 cm of a] (c){\(\phi_{i}\)};
\vertex [right=of a] (d);
\vertex [right=of d] (e) ;
\vertex [above right=of e] (f){\(\phi_{f}\)} ;
\vertex [below right=of e] (g){\(\phi_{f}\)};

\diagram* {
(d) -- [scalar] (e),
(a) -- [scalar] (d),
(b) -- [scalar] (d),
(c) -- [scalar] (d),
(e) -- [scalar] (f),
(e) -- [scalar] (g),
};
\end{feynman}
\end{tikzpicture}

\begin{tikzpicture}
\begin{feynman}
\vertex (a){\(\phi_{i}\)} ;
\vertex [above=0.7 cm of a] (b){\(\phi_{i}\)} ;
\vertex [below=0.7 cm of a] (c){\(\phi_{i}\)};
\vertex [right=of a] (d) ;
\vertex [right=0 cm of d] (e) ;
\vertex [above right=of e] (f) {\(\phi_{f}\)};
\vertex [below right=of e] (g){\(\phi_{f}\)};

\diagram* {
(d) -- [scalar] (e),
(a) -- [scalar] (d),
(b) -- [scalar] (d),
(c) -- [scalar] (d),
(e) -- [scalar] (f),
(e) -- [scalar] (g),
};
\end{feynman}
\end{tikzpicture}

\caption{Feynman diagrams of the $3\to2$ process involving a combination of the $\phi^{3}$ and $\phi^{4}$ vertices (upper diagram) and the only $\phi^{5}$ interactions (lower diagram).}
\label{fig:feynman}
\end{figure}
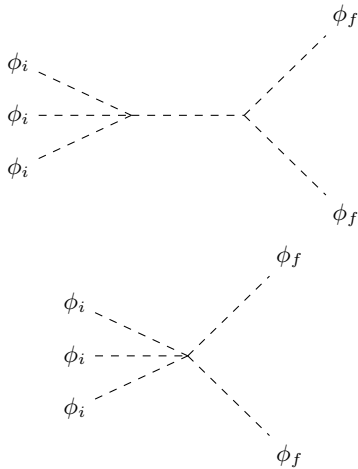

We compared the glueball relic density in Tab.~\ref{tab:relic} with a calculation including only the $\phi^{5}$ interaction, finding a factor $\sim1.3-1.5$ (depending on $N$) of increase in this latter case. This shows that $\phi^{3}$ and $\phi^{4}$ interactions are important in the glueball thermalization process. Indeed, the $3\to2$ number changing process can happen both due to a $\phi^{5}$ order vertex and because of a combination of $\phi^{3}$ and $\phi^{4}$ vertices as shown in Fig.~\ref{fig:feynman}.

Without lower order interactions, just the $ \phi^{5}$ term induces a weaker interaction among the glueballs. Consequently, glueballs freeze-out earlier, when their number is higher and resulting in a larger relic density. In other words, the number-changing interactions have less time to reduce the number of glueballs. This picture is confirmed by the observation that including interactions up to the fourth order ($\phi^{2}$, $\phi^{3}$ and $\phi^{4}$) the relic density increases only of less than $1\%$ compared to the exact calculation involving the log-potential. This reasoning also brings us to the conclusion that higher order number-changing processes, like $4\to2$ interactions, do not have a strong impact on the glueball thermalization. In conclusion, when comparing with results in literature, one should be careful in checking which potential is used. From the perspective of a complete model, not including $\phi^{3}$ and $\phi^{4}$ interactions is inconsistent, leading to a larger relic density. Moreover, note that the form of the glueball potential fixes uniquely, once expanded around the minimum, all the self-interaction couplings at any order. The latter are usually taken to be $\mathcal{O}(1)$ in the literature, while our approach reveals that these numerical coefficients are rather different from 1 and any comparison with the literature must account for this important difference. As a final remark, Eq.~10 in Ref.~\cite{Halverson:2016nfq} is obtained by setting the numerical factors in Eq.~8-9, involving the Lambert W-function, to 1. This is also causing a slight overestimation of the relic density.

The second important point can be understood on the basis of energy considerations. Starting from the effective Lagrangian, it is straightforward to compute the energy density of the gluon field at $T\to \infty$, which corresponds to~\cite{Panero:2009tv,Engels:1999tk}
\begin{equation}
    \rho_{g}=1.21\frac{\pi^{2}}{45}g T^{4}\,,
    \label{eq:rhog}
\end{equation}
with $g=N^{2}-1$ for $SU(N)$. As the temperature approaches the critical one at the phase transition, this energy reduces to match the fitted lattice data. The physical reason is that the dark gluons dissipate energy in the process of bubble nucleation and a smaller energy budget is actually available for the glueball formation. This effect is purely non-perturbative, and it has been taken into account in our analysis once that lattice fits are considered. Calculating the amount of energy stored in dark gluons at the phase transition we realize that it is $3-4$ times smaller than Eq.~\eqref{eq:rhog} depending on $N$. Thus, the energy budget available to glueballs was overestimated if taken equal to the free dark gluon gas approximation. We remark the physical picture behind the energy exchanges within the dark sector. Most of the energy stored in the dark gluon plasma goes into formation of bound states effectively generating the ``mass gap'', i.e. converted into the glueball mass $m_{\rm gb}$. Some of that energy still remains as ``heat'' in the dark sector, essentially in the form of kinetic energy of glueballs, as well as through a contribution to the potential energy from glueball self-interactions. Hence, due to the energy conservation and the absence of interactions with the SM sectors, the totality of energy stored in the dark gluon gas remains in the dark sector. Also, during the phase transition, some part of the energy of the transition (latent heat) is released in the form of gravitational waves~\cite{Reichert:2021cvs,Pasechnik:2023hwv}, a dissipation effect that is neglected in this paper and postponed to a future work.

The aforementioned reduction of the initial energy of the dark gluon gas is partially counterbalanced by a slower dilution of the glueball field compared to a pure cold DM case. Indeed, immediately after the phase transition, the glueball field is rolling fast in a potential which is much larger than its kinetic energy. This results into an equation of state $p=w\rho$ with $-1<w<0$, leading to a slower dilution of the glueball field. Only after this transient phase glueballs act like CDM. To summarize, here we list the various effects explaining the discrepancy with the literature:
\begin{itemize}
    \item we include $\phi^{3}$ and $\phi^{4}$ interactions, instead of only a $\phi^{5}$ vertex. This makes the number changing processes more efficient, reducing the glueball relic density (note that in Ref.~\cite{Forestell:2016qhc} the potential includes  $\phi^{3}$ and $\phi^{4}$ terms, but not $\phi^{5}$);
    \item the glueball potential that we consider has `large' couplings for self-interactions, leading to more than one order of magnitude of suppression in the relic abundance;
    \item part of the energy stored in the dark gluon gas is dissipated in bubble nucleation, reducing the relic abundance of a factor $3-4$;
    \item the slower dilution of the glueball gas compared to CDM goes in the direction of increasing the DM relic abundance.
\end{itemize}
The interplay of all these effects is highly non-trivial, motivating the numerical analysis we developed in this work as a reliable method to compute the glueball DM relic density.

\section{Cosmological histories}
\label{sec:cosmology}

\begin{figure*}[t!]
    \vspace{0.cm}
    \includegraphics[width=0.9\linewidth]{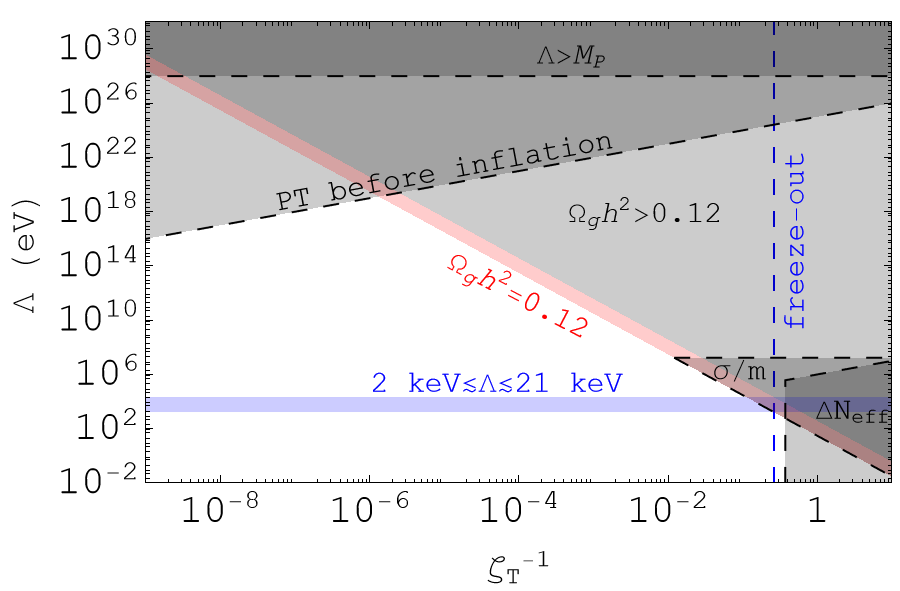}
    \caption{Parameter space of glueball DM expressed in terms of the confinement scale $\Lambda$ and the dark-to-visible sector temperature ratio $\zeta_{T}^{-1}$. Each point in this parameter space represents a possible DM scenario with $\Lambda$ fixed by the properties of the gauge sector and $\zeta_{T}^{-1}$ by its cosmological evolution.
    The gray regions are excluded by various arguments (see the text for more details): the confinement scale has to be sub-Planckian; the phase transition (PT) has to happen after inflation; dark gluons cannot exceed the observed $\Delta N_{\rm eff}$; glueballs cannot overclose the Universe and DM self-interactions are constrained by observations. The red region corresponds to DM fully composed of glueballs; the variability range takes into account the differences between the gauge groups. The blue region for $\zeta_{T}^{-1}\simeq0.26$ is favored in freeze-out models, corresponding to a phase transition scale $2~\keV\lesssim\Lambda\lesssim 21~\keV$ (values corresponding to the largest variability shown in Tab.~\ref{tab:relic}, $30~\eV\lesssim\Lambda_{0}\lesssim 372~\eV$) to have all the DM in the form of glueballs. However, this region is in tension with constraints.}
    \label{fig:paramspace}
\end{figure*}

In order to accurately determine the glueball relic density we must specify the temperature of the dark sector with respect to the photon one; i.~e. the $\zeta_{T}$ parameter. This quantity is vastly unconstrained because of the large number of models predicting different interactions between dark gluons and standard model particles. In a very model-independent fashion, we consider two possibilities for the dark gluon production in the Early Universe: freeze-out and the parent particle decay.

In the first case we assume that, at some point, dark gluons were in thermal equilibrium with the primeval plasma. This is possible due to feeble interactions between the dark and the visible sector. We prefer to keep a model-independent point of view in this work, without specifying the origin of this interaction, but just assuming that it is feeble enough that decoupling happens soon after the end of inflation and the DM is stable on cosmological timescales. A motivated example of feeble interaction is given by fermions that are charged under both the dark and SM gauge groups with a mass much larger than the confining temperature. In this case, all the interactions will be strongly suppressed by the mass scale of this mediator. In the following we do not assume any particular model of interaction and our considerations are  completely model-independent.
When their interaction rate becomes smaller than the Hubble parameter, dark gluons decouple from the thermal bath. In this case the temperature of dark gluons will trace the photon one up to the decoupling, then entropy production events will cause a cooling of the dark sector following
\begin{equation}
    \zeta_{T}^{-1}=\left(\frac{g_{*,s}(T_{\gamma})}{g_{*,s}(T_{d})}\right)^{1/3}\,,
\end{equation}
where $T_{d}$ is the decoupling temperature, which determines the number of entropic degrees of freedom at the freeze-out. Without specifying the interaction between dark and visible sector, we know that the lowest temperature of the dark sector is obtained when $g_{*,s}(T_{d})=106.75$. This corresponds to a weak interaction between the two sectors that leads to a decoupling at $T\gg 100$~GeV, where we assume only standard model particles to exist. This is a strong assumption given our ignorance of physics at very high-energy scales. For instance, in the minimal supersymmetric standard model extension $g_{*,s}(T_{d})=228.75$~\cite{Cadamuro:2011fd}. This latter value will be used to fix the coldest dark gluon scenario, which gives $\zeta_{T}^{-1}\simeq0.26$, where we have used $g_{*,s} (T_{\gamma}) =3.909$.
This consideration gives a sense of how the dark sector can be colder than the visible one, proving that in this scenario of freeze-out the two temperatures are never too different.

On the other extreme, a hot dark gluon sector is constrained by the measurements of the number of relativistic species. Indeed, dark gluons cannot contribute to the effective number of relativistic species $N_{\rm eff}$ more than the constraint $\Delta N_{\rm eff}<0.35$ at the 95$\%$ CL~\cite{Planck:2018vyg}. For a dark sector that is not in equilibrium with the thermal bath, this constraint translates into~\cite{Breitbach:2018ddu}
\begin{equation}
    \Delta N_{\rm eff}=\frac{4}{7}\left(\frac{11}{4}\right)^{4/3}g \zeta_{T}^{-4}\lesssim 0.35\,,
\end{equation}
where the number of degrees of freedom is $g=N^{2}-1$ for $N=\{3,4,5\}$. Therefore, $\zeta_{T}^{-1}\lesssim 0.37$ for $N=3$ and $\zeta_{T}^{-1}\lesssim 0.28$ for $N=5$. We consider the former, more conservative, value as upper limit on $\zeta_{T}^{-1}$. This constraint requires that dark gluons confine after the Big Bang Nucleosynthesis, which happens at $T_{\gamma}\simeq 1$~MeV. 

There are several realizations of the parent particle decay scenario, and we consider the one that we consider the simplest. In this case, a heavy field (that can be associated with the inflaton) decays into standard model particles and also in dark gluons. We take the branching ratio to decay into dark gluons to be $f$ (and $1-f$ is the branching ratio into standard model radiation). Compared to the photon energy density, the dark gluon one decreases because of entropy production events in the visible sector and the temperature evolves accordingly
\begin{equation}
    \zeta_{T}^{-1}=\left(\frac{g_{*,s}(T_{\gamma})}{g_{*,s}(T_{d})}\right)^{1/3}\left(\frac{f}{1-f}\right)^{1/4}\,.
\end{equation}
Compared to the freeze-out case, depending on the value of $f$, the dark sector can be extremely cold compared to the visible one (see also the recent Ref.~\cite{Kolesova:2023yfp} for a discussion on the temperature of confining dark sectors). We assume that the coldest dark sector case corresponds to a confinement happening soon after inflation, when the photon bath has a temperature $T_{\gamma}\simeq 10^{16}$~GeV. This limit corresponds to the breakdown of our calculation, that is valid when the Universe is radiation dominated. This discussion reveals that dark gluons can be extremely cold compared to the thermal photon bath, leading to a very early phase transition, perhaps during inflation. In this case we expect a strong damping of the oscillations of the glueball field, suppressing the relic density. Therefore, we consider that, in order to produce glueball DM, the phase transition cannot happen before or during inflation. This discussion motivates us to use $\zeta_{T}^{-1}$ as a free parameter: if $\zeta_{T}^{-1}\sim\mathcal{O}(1)$, we are modeling a dark sector with interactions strong enough to establish thermal equilibrium soon after inflation; if $\zeta_{T}^{-1}\lesssim 1$, the interactions are so feeble (or even absent) that there is no relation between the temperatures of the visible and dark sectors. Thus, casting this discussion in terms of $\zeta_{T}^{-1}$ is an extremely powerful and general method describing in an unified way several possible models.

We remark that, despite glueballs undergo $3\to 2$ number changing interactions (a cannibalistic phase) for a period of their evolution, this phase has to end before matter-radiation equality in the case that glueballs make up the majority of DM. This condition is verified in the allowed region of the parameter space. When cannibalism is finished, glueballs are mildly relativistic, with an average energy roughly $1.5\,m_{\rm gb}$. After that, they cool down quite rapidly because of the Universe expansion, effectively becoming CDM. As shown in Eq.~\eqref{eq:expansion}, the glueball self-interactions are repulsive and one may wonder if this feature affects structure formation. A simple estimate of this effect is obtained by comparing the intensity of self-interactions, proportional to the glueball field amplitude $\phi_{0}$, with the particle mass $m_{\rm gb}$~\cite{Khlopov:1985jw}. The field amplitude is directly connected to the glueball number density $n_{\rm gb}=\Omega \rho_{c}/m_{\rm gb}\simeq 94\cm^{-3}(\Lambda/\eV)^{-1}$ through $\phi_{0}=\sqrt{n_{\rm gb}/2m_{\rm gb}}\simeq2.5\times10^{-7}\eV (\Lambda/\eV)^{-1}$. Since $m_{\rm gb}\gg \phi_{0}$, the structure formation is indistinguishable from collisionless DM.

This discussion is motivated by the flexibility in modeling interactions between dark gluons and ordinary matter, determining the cosmological evolution of the hidden sector. This results in a broad parameter space available to glueballs to explain the nature of DM.

\section{Conclusions}
\label{sec:conclusions}

In this work we explored in detail the formation of scalar glueballs in generic $SU(N)$ dark gauge sectors. We showed that this composite state is a good DM candidate, viable in several cosmological models. The delicate interplay between microphysics governing the phase transition (the confinement-deconfinement phase transition scale $\Lambda$) and the macroscopic cosmological evolution (the dark-to-visible temperature ratio $\zeta_{T}^{-1}$) determines the relic density of the glueball DM. In Fig.~\ref{fig:paramspace} we summarized our findings in the $\Lambda$ vs $\zeta_{T}^{-1}$ parameter space, showing that a large portion of it is viable and unconstrained. The red band shows the parameter space where glueballs constitute the whole DM, depending on the gauge group considered. Above this line, glueballs would overclose the Universe and this gray region is excluded. Moreover, we require the phase transition to happen in a radiation-dominated era, when the photon bath temperature is assumed to be approximately below $10^{16}$~GeV, otherwise the glueball relic density would be strongly suppressed by inflation. Precisely, we require that the confining scale matches the glueball temperature after inflation, i.e. $T=\Lambda=\zeta_{T}^{-1}T_{\gamma}<\zeta_{T}^{-1}~10^{16}\GeV$.
As another consistency condition, we also mark the region where the confinement scale is super-Planckian.
The blue region shows the range of $\zeta_{T}^{-1}$ easily accommodated in a freeze-out scenario for dark gluons. For example, we find that a window with $2~\keV\lesssim\Lambda\lesssim21$~keV would explain the DM in a simple freeze-out scenario for the dark gluons forming glueballs. However, this region is in tension with
constraints on DM self-interactions, requiring the cross-section to mass ratio be $\sigma/m\lesssim0.19~{\rm cm}^{2}/{\rm g}$~\cite{Eckert:2022qia}. In our model glueballs undergo $2\to2$ scatterings with a cross section $\sigma\sim m_{\Lambda}^{-2}$, in the non-relativistic limit, exceeding the current constraints if $\Lambda\lesssim17$~MeV and glueballs constitute the majority of DM. 

Cosmological observations of $\Delta N_{\rm eff}$ constrain models in which dark gluons are extremely hot, excluding the possibility for a very low confinement scale $\Lambda\ll100$~eV. In conclusion, this study allowed us to delineate a useful parameter space for glueball DM. We expect future studies using observational constraints from cosmology and astrophysics, as well as those produced from upcoming laboratory experiments, will populate Fig.~\ref{fig:paramspace} to provide a clearer picture and better understanding of glueball DM properties and behaviors.

This work sets the necessary basis for future investigations of the glueball phenomenology as a promising candidate for unveiling the dark sector.
Due to its very interdisciplinary nature, a line of research focused on glueballs also opens up the possibility of various synergies between cosmological surveys and collider searches~\cite{CMS:2021dzg,Albouy:2022cin,ATLAS:2023swa}. Besides phenomenological applications, future investigations should focus on comparing the results obtained in the presented formalism with alternative methods to describe the phase transition, in order to assess the robustness of these calculations.

\vspace{0.3cm}
\acknowledgements
P.C. warmly thanks Edward Hardy, Michele Redi, Gustavo Salinas, Juri Smirnov and Andrea Tesi for fruitful discussions.
The work of P.C. is supported by the European Research Council under Grant No.~742104 and by the Swedish Research Council (VR) under grants 2018-03641 and 2019-02337. T.F. is supported by a Royal Society Newton International Fellowship (NIF-R1-221137). R.P.~and Z.-W.~W.~are supported in part by the Swedish Research Council grant, contract number 2016-05996, as well as by the European Research Council (ERC) under the European Union's Horizon 2020 research and innovation programme (grant agreement No. 668679). This article is based upon work from COST Action COSMIC WISPers CA21106, supported by COST (European Cooperation in Science and Technology).

\bibliographystyle{bibi}
\bibliography{biblio.bib}

\end{document}